\documentclass[prb, twocolumn,superscriptaddress]{revtex4}
\usepackage[english]{babel}
\usepackage[dvips]{graphicx}
\usepackage{amssymb}
\usepackage{amsmath}
\begin{document}

\title{Thermoelectrical manipulation of nano-magnets}
\author{A. M. Kadigrobov}
\affiliation{Department of Physics, G{\" o}teborg University, SE-412
96 G{\" o}teborg, Sweden} \affiliation{Theoretische Physik III,
Ruhr-Universit\"{a}t Bochum, D-44801 Bochum, Germany}
\author{R. I. Shekhter}
\affiliation{Department of Physics, G{\" o}teborg University, SE-412
96 G{\" o}teborg, Sweden}
\author{M. Jonson}
\affiliation{Department of Physics, G{\" o}teborg University, SE-412
96 G{\" o}teborg, Sweden} \affiliation{School of Engineering and
Physical Sciences, Heriot-Watt University, Edinburgh EH14 4AS,
Scotland, UK}
\author{V. Korenivski}\affiliation{Nanostructure Physics, Royal Institute
of Technology, SE-106 91 Stockholm, Sweden}
\date{\today}

%\ead{mjonsson@fy.chalmers.se} \address{$\dag$ Department of Applied
%Physics, Chalmers University of Technology and G{\"o}teborg
%University, SE - 412 96 G{\"o}teborg, Sweden}

%\address{$\ddag$ Department of Applied Physics, Chalmers University of
%Technology and G{\"o}teborg University, SE - 412 96 G{\"o}teborg,
%Sweden}

%\pacs{85.35.Kt, 85.85.+j}
\begin{abstract}
We propose a device that can operate as a magneto-resistive switch
or oscillator. The device is based on a spin-thermo-electronic
control of the exchange coupling of two strong ferromagnets through
a weakly ferromagnetic spacer. We show that the local Joule heating
due to a high concentration of current in a magnetic point contact
or a nanopillar can be used to reversibly drive the weak ferromagnet
through its Curie point and thereby exchange-decouple the strongly
ferromagnetic layers, which have an antiparallel ground state. Such
a spin-thermionic parallel-to-antiparallel switching causes
magnetoresistance oscillations where the frequency can be controlled
by proper biasing from essentially DC to GHz.
\end{abstract}
\maketitle

Manipulation of the magnetic state on the nanometer scale is the
central problem of applied magneto-electronics.  The torque effect
\cite{Slonczewski,Berger}, which is based on the exchange
interaction of electrons injected into a ferromagnetic region, is
one of the key phenomena leading to current-induced magnetic
switching. Current-induced precession and switching of the
orientation of magnetic moments due to this effect has been observed
in many experiments
\cite{Tsoi1,Tsoi2,Myers,Katine,Kiselev,Rippard,Beech,Doughton,Prejbeanu,Jianguo,Dieny}.

The efficiency of current-induced switching is, however, limited by
the necessity to work with high current densities. A natural
solution to this problem is to use electrical point contacts (PCs).
This is because now the current density is high only near the PC,
where it can reach\cite{Yanson,Versluijs} values of order
10$^9$~A/cm$^2$. The characteristic energies  supplied to the
electronic system are determined by the voltage drop $V$, and are
for $V\sim 0.1 \div 1$~V (which is easily reached in experiments)
comparable to the exchange energy in magnetic materials.

One of the characteristic features of such high current-density
states is the possibility to create local heating in regions with a
size of a few nanometers. The energy supplied to the electronic
system in this restricted region results in an enhanced local
temperature, which to a high degree of accuracy can be controlled by
the applied voltage. Electrical manipulation of nanomagnetic
conductors by such controlled Joule heating is a new principle for
current-induced magnetic switching. In this Letter we discuss one of
the possibilities of the thermoelectrical magnetic switching caused
by a non-linear interaction between  the spin-dependent electrical
transport and the magnetic sub-system of the conductor due to the
Joule heating effect. We predict that a special design of magnetic
PC can provide both a fast switching and a smooth change of the
magnetization direction in nano-size regions of the magnetic
material controlled by the voltage. We also predict temporal
oscillations of the magnetization direction in such regions (which
are accompanied by electrical oscillations) under an applied DC
voltage. These phenomena are potentially useful for microelectronic
applications such as memory devices and voltage controlled
oscillators.

\begin{figure}
  %%%%%%%%%%%%%%%%%%%%%%%%%%%%%%%% \centerline{\psfig{figure=zlaser2.eps,width=8cm}}
 \centerline{\includegraphics[width=6.0cm]{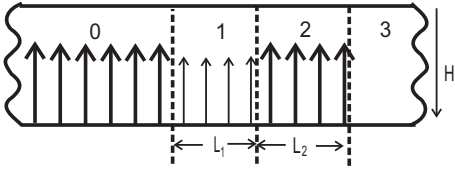}}
  %\vspace{1cm}
  \caption{Orientation of the magnetic moments in
  a  stack of three ferromagnetic layers
(0, 1, 2);  the presence of a magnetic field  directed opposite to
the magnetization is indicated by an arrow outside the stack.}

  \label{noflip}
  \end{figure}

{\it Equilibrium magnetization distribution in a magnetic stack.}
 The system under consideration is  shown in
Fig.\ref{noflip} where three ferromagnetic layers  are coupled to a
non-magnetic conductor. We assume the following conditions to be
satisfied:  The Curie temperature $T_c^{(1)}$ of region 1 is smaller
than the Curie temperatures $T_c^{(0,2)}$ of regions 0 and 2; in
region 2 there is a magnetic field directed opposite to the
magnetization of the region, which can be an external field, the
fringing field from layer 1, or a combination of the two. We require
this combined magnetostatic field to be weak  enough so that at low
temperatures the magnetization of layer 2 is kept parallel to the
magnetization of  layer 0 due to the exchange interaction between
them via region 1 (we assume the magnetization direction of layer 0
to be fixed). In the absence of an external field and if the
temperature $T> T_c^{(1)}$,
 %is higher than the Curie point, the spacer
 the proposed F/f(N)/F tri-layer is similar to the antiparallel spin-flop 'free layers'
widely used in memory applications\cite{Worledge}.

As  $T$  approaches $T_c^{(1)}$  the magnetic moment of layer 1
decreases and the exchange coupling between layers 0 and 2 weakens.
This results in an inhomogeneous  distribution of the stack
magnetization.
 Euler's equation for the magnetization distribution that
 minimizes the free energy of the system  can be written as follows
 (see, e.g., Ref. \onlinecite{Landau}):
\begin{equation}
\frac{d}{d x}\left(\alpha (x)M^2(x)\frac{d\theta}{dx}
\right)-\frac{\beta}{2}M^2
\sin{2\theta}+\frac{HM}{2}\sin{\theta}=0;\label{inhom}
\end{equation}
Here  the $x$- and $z$-axes are directed along the stack and the
magnetization direction in the region 0, respectively; $\theta (x)$
is the angle between the  magnetic moment $\vec M (x)$ at point $x$
and the $z$-axis; $M=|\vec M|$; $\alpha \sim I/a M^2$, where $I\sim
kT_c$ and $a$ is the exchange energy and the lattice spacing,
respectively, $k$ is the Boltzmann constant; $\beta$ is the
dimensionless constant of the anisotropy energy. Below we assume the
lengths $L_{1,2}$ of  regions 1 and 2  to be smaller than the domain
wall lengths in these regions.

%%%%%%%%%%%%%%%%%%%%%%%%%%%%%%%%%%%%%%%%%%%%%%%
  %%%%%%%%%%%%%%%%%%%%%%%%%%%%%%%%%%%%%%%%%%%%%%%
 %%%%%%%%%%%%%%%%%%%%%%%%%%%%%%%%%%%%%%%%%%%%%%%

%%%%%%%%%%%%%%%%%%%%%%%%%%%%%%%%%%%%%%%%%%%%%%%
\begin{figure}
%%%%%%%%%%%%%%%%%%%%%%%%%%%%%%%% \centerline{\psfig{figure=zlaser2.eps,width=8cm}}
 \centerline{\includegraphics[width=6.0cm]{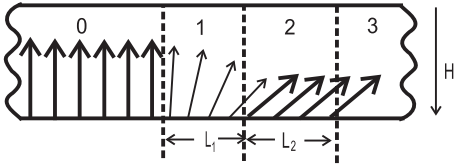}}
%\vspace{1cm}
\caption{Magnetic moment orientations in the stack at temperatures
$T_1<T <T_c^{(1)}$.}
 \label{changeorient}
\end{figure}
%%%%%%%%%%%%%%%%%%%%%%%%%%%%%%%%%%%%%%%%%%%%%%%

Using Eq.~(\ref{inhom}) and the boundary condition at the interface
$S$ between magnetic region 2 and the non-magnetic region 3
$d\theta/dx|_{S}=0$ (see, e.g., Ref.~\onlinecite{Akhiezer})
 one finds the magnetization in
region 1 to be inhomogeneous $\theta_1 (x)=\theta_2 x/L_1,
\hspace{0.25cm}  0\leq x \leq L_1$, while the magnetic moments in
region 2 are approximately parallel with the accuracy $\alpha_1
M_1^2(T)/\alpha_2 M_2(T)\ll 1 $ (the subscripts 1 and 2 refer to the
region 1 and 2, respectively); the dependence of their tilt angle
$\theta_2$ on $H$ and $T$ is given by the equation
\begin{equation}
\theta_2=D(H,T)\sin{\theta_2}; \;\;D(H,T)=\frac{L_1 (L_2H
M_2(T))}{4 \alpha_1  M_1^2(T)}
 \label{theta2}
\end{equation}
 While writing
Eq.(\ref{theta2}) we neglected the anisotropy energy in comparison
with the magnetic energy in region~2.
%$M_1(T)=M_1^{(0)}\sqrt{(T_1^{(c)}-T)/T_1^{(c)}}$ and $M_2(T)$ are
%the magnetic moments of the regions 1 and 2, respectively;
%parameter $D(H,T)$ is a ratio of the magnetic energy  and the
%exchange energy in the stack volume for the inhomogeneous
%distribution of the magnetization shown in Fig.\ref{changeorient})

%One easily sees that inside the interval $0\leq \theta_2 \leq \pi$
%Eq.(\ref{theta2}) has either one  or two roots . As these roots
%correspond to extrema of the free energy, it is obvious that in
%the former case it has one minimum and in the latter case - one
%minimum and one maximum.
At low temperatures ($D(T,H) <1$) Eq.(\ref{theta2}) has the single
root $\theta_2=0$ and hence a parallel orientation of all the
magnetic moments in the stack is thermodynamically stable. At
temperatures $T\geq T_1$ (here $T_1$ is the temperature at which
$D(T,H)=1$)  one has $D\geq 1$ and a second root $\theta_2 \neq 0$
appears in addition to the root $\theta_2=0$. The parallel
magnetization corresponding to $\theta_2=0$ is now unstable and the
magnetization direction of region 2  tilts with an increase of $T$
(see Fig.\ref{changeorient}) until at $T \geq T_c^{(1)}$ the
 exchange coupling between layers 0 and 2 vanishes and their
 magnetic moments become antiparallel.

 {\it Current-voltage characteristics of the stack under Joule
heating.} If the stack is Joule heated  by the current $J$ its
temperature $T(V)$ is determined by  the heat-balance condition
\begin{equation}
JV=Q(T), \hspace{0.2cm}J  =V/R(\theta_2)
 \label{heat}
\end{equation}
and Eq.~(\ref{theta2}) that determines the temperature dependence of
$\theta_2(T(V))$. Here $Q(T)$ is the heat flux from the stack,
$R(\theta_2)$ is the stack resistance. Eqs.~(\ref{theta2}) and
(\ref{heat}) define the current-voltage characteristics (IVC) of the
stack $J=R(\theta_2(V))V$ in a parametric form.

\begin{figure}
%%%%%%%%%%%%%%%%%%%%%%%%%%%%%%%% \centerline{\psfig{figure=zlaser2.eps,width=8cm}}
 \centerline{\includegraphics[width=6.0cm]{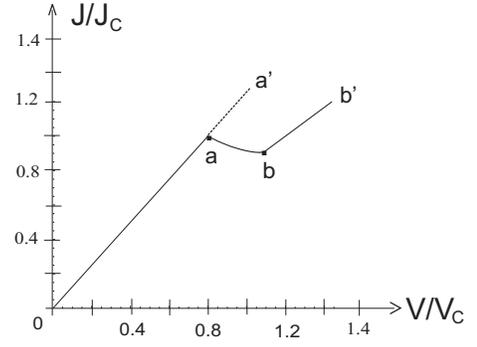}}
\vspace{1cm} \caption{  Numerical results for the $I-V$
characteristics (IVC) of the magnetic stack. Here
$R(\theta_2)=R_{+}-R_{-} \cos{\theta_2}$, $R_-/R_+ =0.2$, $D_0=0.2$;
$J_c=V_c/R_{+}$. The branches  $0-a'$ and $b-b'$ of the IVC
correspond to the parallel and antiparallel orientations of the
stack magnetization, respectively; the branch $a-b$ corresponds to
the inhomogeneous magnetization distribution shown in
Fig.\ref{changeorient}
% the unstable IVC
%branch corresponding to $\theta_2=0$  at Joule heating $T(V)>T_1$
%is shown with a dashed line.
}
 \label{cvc}
\end{figure}

As follows from the previous paragraph, in the whole intervals
$T(V) < T_1$ and $T(V)>T_c^{(1)}$ the stack resistances are $R(0)$
and $R(\pi)$, respectively, that is
 IVC branches $J=R(0)V$ and $J=R(\pi)V$ are linear at $V< V_{1}=\sqrt{R(0)Q(T_1)}$
and $V > V_{c}=\sqrt{R(\pi)Q(T_c^{(1)})}$, respectively. At $V_1
\leq V \leq V_c$ the stack temperature is  $T_1 \leq T(V) \leq
T_c^{(1)}$, and the magnetization direction of region 2 changes
with a change of $V$ and hence  IVC  is non-linear there.
Differentiating Eqs.(\ref{theta2},\ref{heat}) with respect to $V$
one finds that inside this interval  the differential conductance
$R_{d}(V) = d J/dV$ is negative if
$d[G(\theta_2)(1-D_0\sin{\theta_2}/\theta_2)]/d\theta_2<0$ (here
$D_0 =D(H,T=0)$); e.g., for the stack resistance of the form
$R(\theta_2)= R_{+} -R_{-} \cos{\theta_2})$ (here
$R_{\pm}=(R(\pi)\pm R(0))/2$) one has $dJ/dV < 0$  if $D_0(H) <3
r/(1+3r)$  where $r=R_{-}/R_{+}$, and hence IVC of the stack is
N-shaped as shown in Fig.\ref{cvc}.

\begin{figure}
  %%%%%%%%%%%%%%%%%%%%%%%%%%%%%%%% \centerline{\psfig{figure=zlaser2.eps,width=8cm}}
 \centerline{\includegraphics[width=6.0cm]{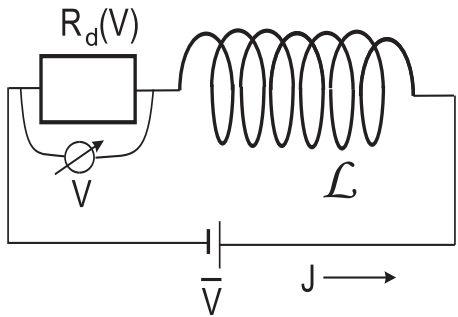}}
  %\vspace{0.2cm}
  \caption{Equivalent circuit for a Joule-heated magnetic stack.
  A differential resistance $R_d(V)$ biased by a fixed DC voltage $\bar V$ is connected in series with an
  inductance ${\cal L}$; $V(t)$ is the voltage drop over the
   stack and $J(t)$ is the total current}

  \label{cvceffcirc}
  \end{figure}

{\it Self-exciting electric, thermal and magnetic direction
oscillations}.  The set of equations describing thermal and
electrical processes in the stack connected  in series with an
inductance  ${\cal L}$ and DC voltage drop $\bar V$ (see
Fig.\ref{cvceffcirc}) is
\begin{equation}
C_V\frac{dT}{dt}=J^2 R(\theta_2)-Q(T); \;\ \;
%\nonumber \\
{\cal L}\frac{dJ}{dt} +JR(\theta_2)={\bar V} %\hspace{1.7cm}
 \label{evolution}
\end{equation}
where $C_V$ is the heat capacity. For the case that the magnetic
moment relaxation is the fastest process the temperature
dependence of $\theta_2=\theta_2(T(t))$ is given by
Eq.(\ref{theta2}).

%%%%%%%%%%%%%%%%%%%%%%%%%%%%%%%%%%%%%%%%%%%%%
  %%%%%%%%%%%%%%%%%%%%%%%%%%%%%%%%%%%%%%%%%%%%%%%
 %%%%%%%%%%%%%%%%%%%%%%%%%%%%%%%%%%%%%%%%%%%%%%%

\begin{figure}
  %%%%%%%%%%%%%%%%%%%%%%%%%%%%%%%% \centerline{\psfig{figure=zlaser2.eps,width=8cm}}
  \centerline{\includegraphics[width=6.0cm]{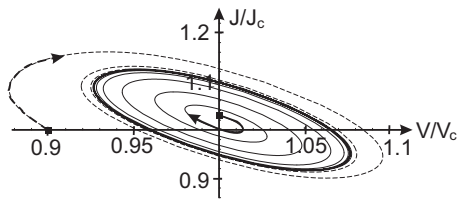}}
  %\vspace{1cm}
  \caption{Numerical results showing spontaneous oscillations of the current $J(t)$
  and the voltage drop $V(t)$ over the stack for the parameters
  $R_-/R_+=0.2$,  $D_0=0.1$ and
$({\cal L} -{\cal L}_{cr})/{\cal L}_{cr}=0.25$; $J_c=V_c/R_+$. The
time development of the current and the voltage drop over the stack
follows the dashed line or the thin solid line towards
  the limiting cycle (thick solid line) depending on the initial state.}
   \label{cvclim1}
  \end{figure}

The set of equation Eq.(\ref{evolution}) has always a steady-state
solution Eqs.(\ref{theta2},\ref{heat}). A study of its stability
with respect to small perturbations shows that inside the interval
$V_1 \leq {\bar V} \leq V_c$ this solution is unstable if ${\cal
L}>{\cal L}_{cr}= t_v|R_d({\bar V})|$ where $t_v \sim (T C_V
R_+/{\bar V}^2)$ is the characteristic time of the voltage
evolution and $R_d =d V/dJ <0$ is the stack differential
resistance. As a result periodic oscillations of the current
$J(t)$ and the voltage drop on the stack $V(t)$ spontaneous arise:
the current $J(t)$ and the voltage drop on the stack $V(t)$
periodically move along the limiting cycle in the plane $(J,V)$
 as is shown in Fig.\ref{cvclim1}. The temperature of the stack $T(t)$
 and the magnetization direction $\theta_2(t)$ adiabatically
 follows these electric oscillations according to the following
 equalities: $Q(T(t))= V(t)J(t)$ and $\theta_2(t)=\theta_2(T(t))$
 (see Eq.(\ref{theta2}))
%%%%%%%%%%%%%%%%%%%%%%%%%%%%%%%%%%%%%%%%%%%%%%%
  %%%%%%%%%%%%%%%%%%%%%%%%%%%%%%%%%%%%%%%%%%%%%%%
 %%%%%%%%%%%%%%%%%%%%%%%%%%%%%%%%%%%%%%%%%%%%%%%
For ${(\cal L}-{\cal L}_{cr})/{\cal L}_{cr}\ll1$ the oscillation
frequency is $\omega =1/t_v$.

With a further increase of the inductance the character of the
periodic motion changes its character: at ${\cal L}\gg {\cal
L}_{cr}$ the current and the voltage slowly move along the
branches of the IVC $0a$ and $bb^{'}$ with the velocity ${\dot
J}/J \approx 1/t_j$ (here $t_J ={\cal L}/R_+$ is the
characteristic time of the current evolution) fast switching
between them at points $a$ and $b$ with the velocity $\sim 1/t_v$.
Therefore, in this case the stack periodically switches between
the parallel and  antiparallel magnetic states.

{\it In conclusion}, we have shown that Joule heating of the
magnetic stack presented in Fig.\ref{noflip} allows electrical
manipulations of the mutual magnetization orientation   of layers
0 and 2 as follows:

1) In the regime of a controlled  current there is a hysteresis
loop: with an increase of the current along the a-a' IVC branch
the magnetic directions are parallel up to point $a$. At this
point the stack is heated up to the temperature $T_1$ at which
$D=1$ and the parallel oriented state looses its stability. As a
result the system switches to the $b-b'$ branch of  IVC   at which
$T>T_c^{(1)}$ and the magnetic moment $\vec{M}_2$ switches to the
antiparallel orientation. With a further change of the current the
stack remains on the b-b' branch of  IVC until the current reaches
point $b$ where the stack is cooled below $T_c^{(1)}$ and system
switches again to the $a-a'$ branch at which the magnetic moments
are parallel.

2) At the regime of a controlled voltage and a low inductance one
may smoothly re-orient the magnetization directions from the
parallel
 to the antiparallel
(see Fig.\ref{changeorient})  changing the applied voltage inside
the interval $V_1 \leq V\leq V_c$ that is moving along $a-b$
branch of IVC.

3) An increase of the  inductance of the circuit under a fixed DC
voltage (see Fig.\ref{cvceffcirc}) allows  to excite electrical
periodic oscillations accompanied by periodical switchings of the
magnetization directions in the stack from parallel to the
antiparallel orientation.

{\it Acknowledgement.} Financial support from the Swedish KVA, VR,
and SSF is gratefully acknowledged.

\end{document}